\documentclass[twocolumn,superscriptaddress,showpacs,pra]{revtex4}
\usepackage{epsfig}
\usepackage{color}
\usepackage{delarray}
\usepackage{amsmath, amssymb}
\usepackage{bm}
\usepackage{float}

\begin{document}
\title{Compressive spectral method for the simulation of the water waves}

\author{Cihan Bay\i nd\i r}
\email{cihan.bayindir@isikun.edu.tr}
\affiliation{Department of Civil Engineering, I\c s\i k University,  \.{I}stanbul, Turkey}

%\date{\today}
\begin{abstract}

In this paper an approach for decreasing the computational effort required for the spectral simulations of the water waves is introduced. Signals with majority of the components zero, are known as the sparse signals. Like majority of the signals in the nature it can be realized that water waves are sparse either in time or in the frequency domain.  Using the sparsity property of the water waves in the time or in the frequency domain, the compressive sampling algorithm can be used as a tool for improving the performance of the spectral simulation of the water waves. The methodology offered in this paper depends on the idea of using a smaller number of spectral components compared to the classical spectral method with a high number of components. After performing the time integration with a smaller number of spectral components and using the compressive sampling technique, it is shown that the water wave field can be reconstructed with a significantly better efficiency compared to the classical spectral method with a high number of spectral components, especially for long time evolutions.

For the sparse water wave model in the time domain the well-known solitary wave solutions of the Korteweg-deVries (KdV) equation is considered. For the sparse water wave model in the frequency domain the well-known Airy (linear) ocean waves with Jonswap spectrum is considered. Utilizing a spectral method, it is shown that by using a smaller number of spectral components compared to the classical spectral method with a high number of components, it is possible to simulate the sparse water waves with negligible error in accuracy and a great efficiency especially for large time evolutions.

\end{abstract}
\maketitle

%%%%%%%%%%%%%%%%%%%%%%%%%%%%%%% main %%%%%%%%%%%%%%%%%%%%%%%%%%%%%
%\begin{section}{Introduction}
\section{Introduction}
The signals, with majority of the components are zero, are called sparse signals. Like majority of the signals in the nature, the water waves are sparse either in time or in the frequency domain. Therefore compressive sampling can be thought as a very efficient tool for measuring or simulating the ocean waves. In this paper it is shown that the efficiency of the compressive sampling technique can also be used for the improvement of the computational simulations of the sparse water waves. In coastal engineering literature various types of solitary wave forms are used in tsunami and ocean wave run-up modeling. Therefore for the sparse water wave model in the time domain the solitary wave solutions of the Korteweg-deVries (KdV) equation is considered, which are in the form of the function $sech^2$. For the sparse wave model in the frequency domain, the Jonswap ocean wave spectrum which is frequently used for the representation of the ocean waves is considered.

The water waves are simulated by implementation of a periodic spectral method in which the spectral derivatives are evaluated using the fast Fourier transforms (FFT). Time integration is carried out using a $4^{th}$ order Runge-Kutta method. It is shown that by using a smaller number of spectral components and the compressive sampling technique, water waves can be simulated very efficiently compared to the classical spectral simulation with a higher number of spectral components. Also it is shown that the accuracy difference between two models is negligible.

\section{Methodology}

\subsection{Review of the Solitary Waves and Fourier Spectral Method}

In wave theory there are various forms of the KdV equation which are modified to account for different nonlinear and dispersion effects of the wave field. In this study one of the most early forms of the KdV equation is considered which is \cite{lamb}
\begin{equation}
 \eta_t+\eta \eta_x+\eta_{xxx}=0
\label{eq1}
\end{equation}
where $\eta$ denotes the water surface fluctuation. This equation can be integrated to obtain the well-known solitary wave solutions of the form
\begin{equation}
 \eta(x,t)=3A \textnormal{sech}^2\left(\frac{\sqrt{ A}}{2}(x-At) \right)
\label{eq2}
\end{equation}
where $A$ is a constant. This solitary wave profile can be described as an initial condition for a numerical scheme which solves the KdV equation. One of the most widely used techniques is the periodic spectral method in which the spectral derivatives are evaluated by utilizing efficient FFT algorithm \cite{bay2015a, bayindir2015arxivchbloc, bay2015b, bay2015c, Karjadi2010, Karjadi2012, trefethen}. Numerical time integration is generally performed by schemes such as $4^{th}$ order Runge-Kutta or Adams-Bashforth \cite{bay2009, bay2015d, bay2015e, bay2016, canuto}. The computational spectral method summarized below is given in \cite{trefethen}. 

It is possible to rewrite the KdV equation as
\begin{equation}
 \eta_t+\frac{1}{2}(\eta^2)_x+\eta_{xxx}=0
\label{eq3}
\end{equation}
Taking the Fourier transform of this equation leads to 
\begin{equation}
 \widehat{ \eta}_t+\frac{i}{2}k(\widehat{ \eta^2})-ik^3\widehat{ \eta}=0
\label{eq4}
\end{equation} 
where
\begin{equation}
 \widehat{ \eta}(k,t)=\int^\infty_{-\infty} e^{-ikx}\eta(x,t)dx
\label{eq5}
\end{equation} 
is the Fourier transform of the $\eta$. Multiplying (\ref{eq3}) with an integrating factor of $e^{-ik^3t}$ in order to avoid high wavenumber problems caused by the third term in the equation (stiff term) and by defining $ \widehat{v}=e^{-ik^3t} \widehat{\eta}$, (\ref{eq3}) can be rewritten as
\begin{equation}
 \widehat{v}_t+\frac{i}{2}e^{-ik^3t}k(\widehat{ \eta^2})=0
\label{eq6}
\end{equation} 
which leads to
\begin{equation}
 \widehat{v}_t+\frac{i}{2}e^{-ik^3t}kF\left( \left(  F^{-1} \left(e^{ik^3t} \widehat{v} \right)  \right)^2 \right)=0
\label{eq7}
\end{equation} 
where $F$ and $F^{-1}$ denote the Fourier and inverse Fourier transformations respectively. This equation is solved by a $4^{th}$ order Runge-Kutta method for time integration in order to simulate the solitary waves. Details and further discussion can be seen in \cite{bay2009, demiray, trefethen}.

\subsection{Review of a Linear Ocean Wave Model}
Many different approximate equations are developed to model the ocean waves. Also there are various models which solve the fully nonlinear kinematic and dynamic boundary conditions \cite{bay2009}. In order to discuss the efficiency of the proposed technique, only well-known linear ocean waves are considered in this study. Linearized kinematic and dynamic boundary conditions for the ocean waves at $z=0$ are given as 
\begin{equation}
 \eta_t-\varphi_z=0  \ \ \ \ \ \ \ \ \ \ \ \ \ \ \ \ \varphi_t +g\eta=0
\label{eq8}
\end{equation} 
where $\eta$ and $\varphi$ denote the water surface fluctuation and the velocity potential, respectively \cite{bay2009}. Although this set of equations can be solved analytically to yield sinusoidal waveforms, in order to discuss the advantages of the compressive sampling technique, a numerical spectral method is implemented for simulating the linear ocean waves. In a periodic domain with arbitrary depth $h$, the velocity potential $\varphi$ can be expressed by 
\begin{equation}
\varphi(\vec{x},z,t)=\sum_{n}^{\infty} \varphi_n(t) \frac{\cosh \left[\left|\vec{k_n}\right|(z+h) \right]}{\cosh \left[\left| \vec{k_n} \right| h \right]} e^{i\vec{k_n} \cdot \vec{x}}
\label{eq10}
\end{equation}
where $\vec{k_n}$ is the wave-number vector \cite{bay2009,Karjadi2010,Karjadi2012}. $\varphi_z$ at $z=0$ can be written as
\begin{equation}
\varphi_z(\vec{x},0,t) =F^{-1} \left( F (\varphi) \left| \vec{k_n} \right|  \tanh \left[\left| \vec{k_n} \right| h \right] \right) 
\label{eq10}
\end{equation}
where $F$ and $F^{-1}$ denote the Fourier and inverse Fourier transformations respectively. In this paper only one dimensional waves are studied.

This set of equations are solved with a $4^{th}$ order Runge-Kutta method for time integration in order to simulate the linear ocean waves for various initial linear ocean wave profiles. Initial linear ocean wave fields are constructed by an inverse FFT algorithm using random spectral components with total energy described by the Jonswap spectrum. Details and further discussion can be seen in \cite{bay2009}.

\subsection{Review of the Compressive Sampling}

Since it has been introduced to the scientific community by \cite{candes}, compressive sampling (CS) has drawn the attention of many researchers. Currently it is a common tool in various branches of applied mathematics and some studies such as the development of a single pixel camera system aim to make use of this efficient technique in digital sensing hardware systems as well. In this section a brief summary of the CS is given.

Let $\eta$ be a $K$-sparse signal of length $N$, that is only $K$ out of $N$ elements of the signal are nonzero. $\eta$ can be represented using a orthonormal basis functions with transformation matrix ${\bf \Psi}$. Typical transformation used in literature are Fourier, discrete cosine or wavelet transforms just to mention few. Therefore one can write $\eta= {\bf \Psi} \widehat{ \eta}$ where $\widehat{ \eta}$ is the transformation coefficient vector. Since $\eta$ is a $K$-sparse signal one can discard the zero coefficients and obtain $\eta_s= {\bf \Psi}\widehat{ \eta}_s$  where $\eta_s$ is the signal with non-zero elements only.

The idea underlying in the CS is that a $K$-sparse signal $\eta$ of length $N$ can exactly be reconstructed from $M \geq C \mu^2 ({\bf \Phi},{\bf \Psi}) K \textnormal{ log (N)}$ measurements with an overwhelmingly high probability, where $C$ is a positive constant and $\mu^2 (\Phi,\Psi)$ is coherence between the sensing basis ${\bf \Phi}$ and transform basis ${\bf \Psi}$ \cite{candes}.

Taking $M$ random projections by using the sensing matrix ${\bf \Phi}$ one can write  $g={\bf \Phi} \eta$. Therefore the problem can be recognized as
\begin{equation}
\textnormal{ min} \left\| \widehat{ \eta} \right\|_{l_1}   \ \ \ \  \textnormal{under constraint}  \ \ \ \ g={\bf \Phi} {\bf \Psi} \widehat{ \eta}
\label{eq14}
\end{equation}
where $\left\| \widehat{ \eta} \right\|_{l_1}=\sum_i \left| \widehat{ \eta}_i\right|$. So that among all signal which satisfies the given constraints, the ${l_1}$ solution of the CS problem can be given as  $\eta_{{}_{CS}} ={\bf \Psi} \widehat{ \eta}$. 

$l_1 $ minimization is only one of the alternatives which can be used for this optimization problem. There are some other algorithms to recover the sparse solutions such as greedy pursuit, reweighted $l_1 $ minimization algorithms. Details of this derivation and CS can be seen in \cite{candes}.

\subsection{Proposed Methodology}
%In this study a methodology which can significantly reduce the computational effort required for the spectral simulations of the sparse water waves is offered. 
The methodology summarized here is first described in \cite{bay2015a, bayindir2015arxivcssfm}. In a classical spectral method let $N$ be the number of the spectral components used for representation of a signal. By using only $M$ spectral components with $M << N$ and together with the CS technique to construct the $N$-component signal at the last step of the time evolution, it is possible to obtain a very efficient computational method especially for very long time evolutions. This method can be named as compressive spectral method. Depending on the width of the $K$-sparse wave profile, the selection of the number $M$ has to be done carefully to satisfy $M=O(K \log({N/K}))$ condition of the compressive sampling algorithm. Starting from the initial conditions, time integration is performed for only $M$ spectral components. At the end of the time stepping, the $N$ point signal is reconstructed from $M$ points with the help of the $l_1$ minimization technique of the compressive sampling theory. It is shown that compared to the classical spectral method with $N$ components, the methodology offered in here can reduce the computational effort significantly while the accuracy difference in the results is negligible. The sparse wave models considered in this study are the solitary waves in the time domain and the Airy (linear) ocean waves with total energy described by Jonswap spectrum in the frequency domain.

\section{Results and Discussion}
\subsection{Results for the Time Domain Solitary Wave Simulations}In the Figure~\ref{figone} below, the water surface fluctuations of the classical $N=1024$ component spectral
 
\begin{figure}[h]
\begin{center}
   \includegraphics[width=3.4in]{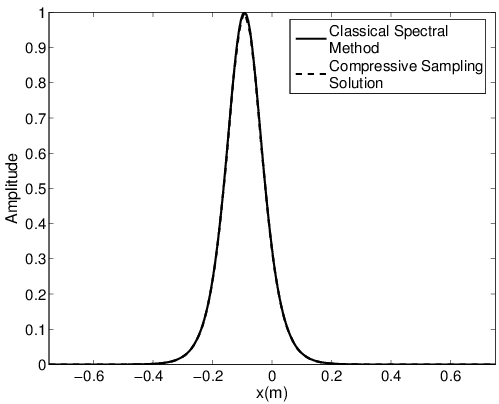}
  \end{center}
\caption{Comparison of the classical spectral method and the proposed compressive spectral method for a solitary wave with $N=1024,M=256$. \label{figone}}
\end{figure}
\noindent method  and the $M=256$ component compressive spectral method proposed is compared for a profile with only one solitary wave. The two methods are in excellent agreement as it can be seen in the figure. The root-mean-square (all rms calculations in this paper are normalized) difference between two profiles is $0.0011$ for this simulation.

\begin{figure}[h]
\begin{center}
   \includegraphics[width=3.4in]{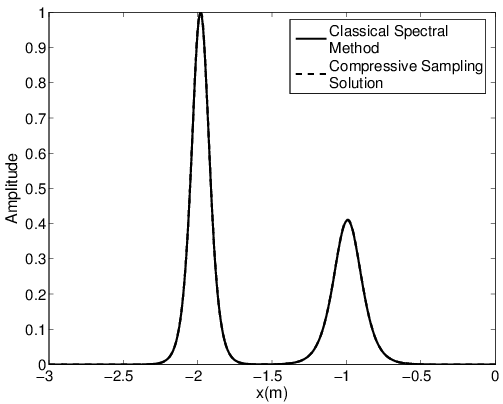}
  \end{center}
\caption{Comparison of the classical spectral method and the proposed compressive spectral method for two solitary waves with $N=2048,M=256$. \label{figtwo}}
\end{figure}

In the Figure~\ref{figtwo} below, the water surface fluctuations of the classical $N=2048$ component spectral method and the $M=256$ component compressive spectral method proposed is compared for a profile with two solitary waves. The two methods are in excellent agreement as it can be seen in the figure. The root-mean-square difference between two profiles is $0.0007$ for this simulation. 

The water surface fluctuations of the classical $N=1024$ component spectral method and the $M=128$ component compressive spectral method proposed is compared for a profile with three solitary waves in the Figure~\ref{figthree} below. The two methods are in excellent agreement as it can be seen in the figure. The root-mean-square difference between two profiles is $0.0019$ for this simulation.

\begin{figure}[h]
\begin{center}
   \includegraphics[width=3.4in]{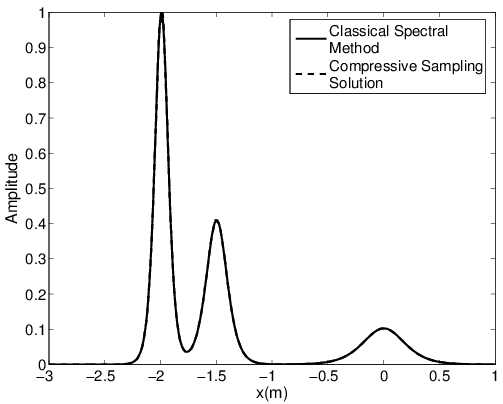}
  \end{center}
\caption{Comparison of the classical spectral method and the proposed compressive spectral method for three solitary waves with $N=1024,M=128$. \label{figthree}}
\end{figure}

All of the results presented above show promising evidence for the accuracy of the proposed method. Additionally the computational efforts required to run the various scenarios are summarized in the Table~\ref{tabone} below. The average computation times of 50 realizations given in the table are in the units of seconds. The times are measured on a Dell Vostro 1700 laptop with dual core of 1.8 GHz and 1GB RAM which is used to run the MATLAB code. As it can be seen on the table, for a very small number of time steps the compressive spectral method does not provide any improvement in the computational effort. This is mainly due to the computational effort required by the ${l_1}$ minimization. However as the number of time steps gets bigger, the computational effort significantly reduces while the differences in the wave profiles are of negligible importance. Therefore compressive spectral method provides a great computational efficiency compared to the classical spectral method.

\begin{table}[H]
\begin{center}
\caption{Comparison of Temporal Cost of the Classical Spectral vs Proposed Compressive Spectral Method-Solitary Wave Simulations.\label{tabone}}
\vspace{10pt}
%\tbl{Comparison of Temporal Cost of the Classical Spectral vs Proposed Compressive Spectral Method.\label{tabone}}
{\begin{tabular}{@{}ccccccc@{}} \toprule
$N$ & $M$ & T. Steps & SSFM-T. (s) & CSSFM-T. (s) & Rms Diff.  \\
\hline
1024\hphantom{00} & \hphantom{0}128 & \hphantom{0}20000& 116.98 & 38.71 &  0.0140 \\
1024\hphantom{00} & \hphantom{0}256 & \hphantom{0}20000& 161.34 & 135.38 & 0.0011 \\
1024\hphantom{00} & \hphantom{0}256 & \hphantom{0}60000& 366.88 & 152.26 & 0.0037 \\
1024\hphantom{00} & \hphantom{0}256 & \hphantom{0}100000& 618.92 & 132.13 & 0.0019 \\
2048\hphantom{00} & \hphantom{0}256 & \hphantom{0}50000& 705.15 & 601.74 & 0.0077 \\
\hline
%0.1\hphantom{00} & \hphantom{0}876.0 & \hphantom{0}875.74 & 0.03 \\
%0.01\hphantom{0} & 2441.0 & 2441.0\hphantom{0} & 0.0\hphantom{0} \\
%0.001 & 4130.0 & 4129.3\hphantom{0} & 0.16\\ \botrule
\end{tabular} }
%\begin{tabnote}
\end{center}
\end{table}
%\end{doublespace}

\subsection{Results for Linear Ocean Wave Simulations}
In the Figure~\ref{figfour} below, the Jonswap spectra of the classical $N=1024$ component spectral method and the $M=256$ component compressive spectral method proposed are compared. The two methods are in excellent agreement as it can be seen in the figure. The root-mean-square difference between two spectra is $0.0045$ for this simulation. 

\begin{figure}[h!]
\begin{center}
   \includegraphics[width=3.4in]{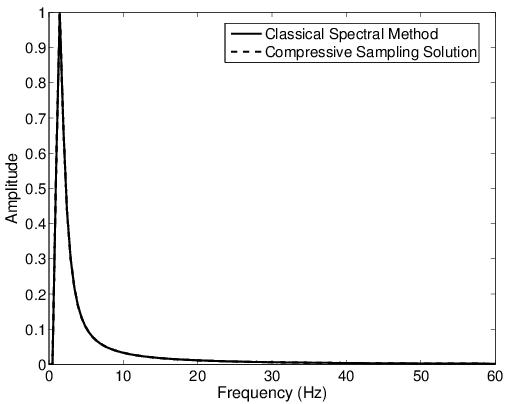}
  \end{center}
\caption{Comparison of the energy spectra of the classical spectral method and the proposed compressive spectral method for linear waves with $N=1024,M=256$. \label{figfour}}
\end{figure}

By means of an inverse FFT, it is possible to construct the ocean surface with linear waves which is not necessarily sparse. In the Figure~\ref{figfive} below, the water surface fluctuations of the classical $N=1024$ component spectral method and the $M=256$ component compressive spectral method proposed are compared. The two methods are in excellent agreement as it can be seen in the figure. The root-mean-square difference between two profiles is $0.0052$ for this simulation. 

\begin{figure}[h!]
\begin{center}
   \includegraphics[width=3.4in]{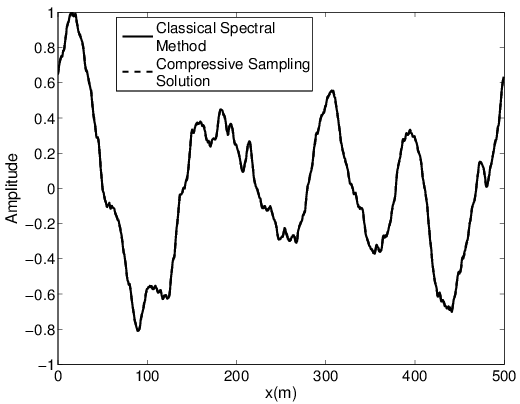}
  \end{center}
\caption{Comparison of the water surface fluctuation of the classical spectral method and the proposed compressive spectral method for linear waves with $N=1024,M=256$. \label{figfive}}
\end{figure}

In the Figure~\ref{figsix} below, the Jonswap spectra of the classical $N=2048$ component spectral method and the $M=256$ component compressive spectral method proposed are compared. The two methods are in excellent agreement as it can be seen in the figure. The root-mean-square difference between two spectra is $0.0044$ for this simulation. 

\begin{figure}[h!]
\begin{center}
   \includegraphics[width=3.4in]{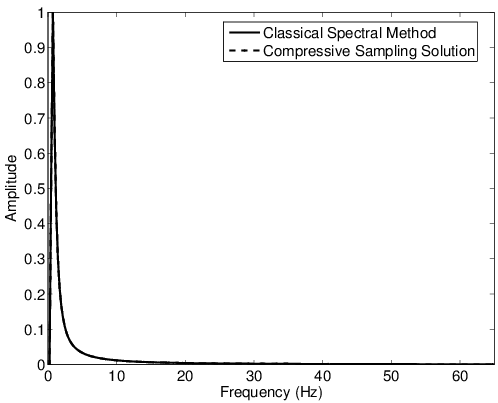}
  \end{center}
\caption{Comparison of the energy spectra of the classical spectral method and the proposed compressive spectral method for linear waves with $N=2048,M=256$.   \label{figsix}}
\end{figure}

Again by means of an inverse FFT, it is possible to construct the ocean surface with linear waves which is not necessarily sparse. In the Figure~\ref{figseven} below, the water surface fluctuations of the classical $N=1024$ component spectral method and the $M=256$ component compressive spectral method proposed are compared. Again the two methods are in excellent agreement in terms of accuracy. The root-mean-square difference between two profiles is $0.0048$ for this simulation. 

\begin{figure}[h!]
\begin{center}
   \includegraphics[width=3.4in]{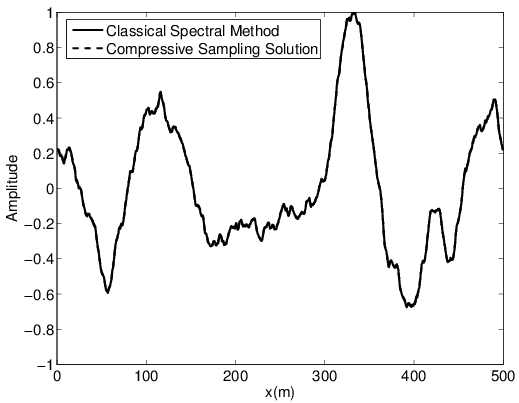}
  \end{center}
\caption{Comparison of the water surface fluctuation of the classical spectral method and the proposed compressive spectral method for linear waves with $N=2048,M=256$. \label{figseven}}
\end{figure}

The Jonswap spectra of the classical $N=1024$ component spectral method and the $M=128$ component compressive spectral method proposed are compared in the Figure~\ref{figeight} below. The results of the two methods agrees very good however it can be realized that using a smaller $M$ causes the difference to increase although still it is of negligible importance. This is mainly due to the fact that for a smaller $M$ the sparsity condition becomes critical. The root-mean-square difference between two spectra is $0.0036$ for this simulation.

\begin{figure}[h!]
\begin{center}
   \includegraphics[width=3.4in]{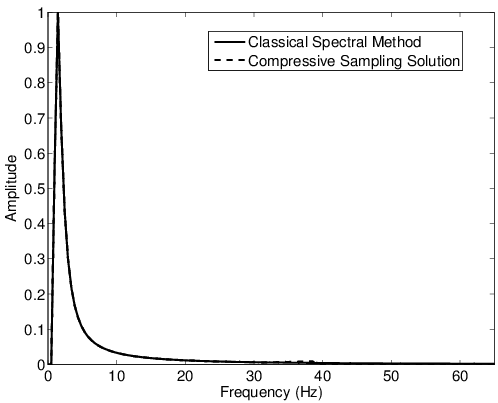}
  \end{center}
\caption{Comparison of the energy spectra of the classical spectral method and the proposed compressive spectral method for linear waves with $N=1024,M=128$.\label{figeight}}
\end{figure}

Again by means of an inverse FFT, it is possible to construct the ocean surface with linear waves which is not necessarily sparse. In the Figure~\ref{fignine} below, the water surface fluctuations of the classical $N=1024$ component spectral method and the $M=256$ component compressive spectral method proposed are compared. The two methods are in very good agreement as it can be seen in the figure. The root-mean-square difference between two profiles is $0.0165$ for this simulation. 

All of the results presented above show promising evidence for the accuracy of the proposed method not only for solitary wave simulations but also linear ocean wave simulations. Additionally the computational efforts required to run the various configurations for the linear ocean wave simulations are summarized in the Table~\ref{tabtwo} below. 

\begin{figure}[h!]
\begin{center}
   \includegraphics[width=3.4in]{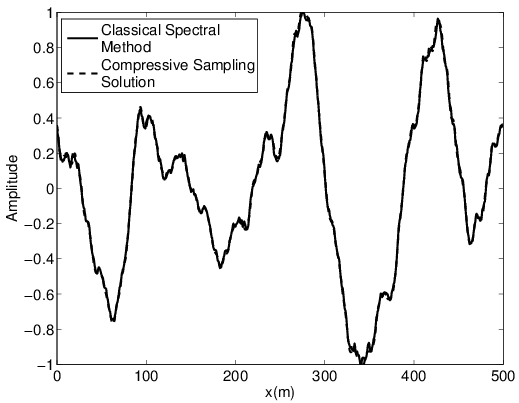}
  \end{center}
\caption{Comparison of the water surface fluctuation of the classical spectral method and the proposed compressive spectral method for linear waves with $N=1024,M=128$.\label{fignine}}
\end{figure}

\begin{table}[H]
\begin{center}
\caption{Comparison of Temporal Cost of the Classical Spectral vs Proposed Method-Linear Wave Simulations.\label{tabtwo}}
\vspace{10pt}
%\tbl{Comparison of Temporal Cost of the Classical Spectral vs Proposed Compressive Spectral Method.\label{tabone}}
{\begin{tabular}{@{}ccccccc@{}} \toprule
$N$ & $M$ & T. Steps & SSFM-T. (s) & CSSFM-T. (s) & Rms Diff.  \\
\hline
1024\hphantom{00} & \hphantom{0}128 & \hphantom{0}30000& 61.90 & 8.91 &  0.0036 \\
1024\hphantom{00} & \hphantom{0}256 & \hphantom{0}70000& 146.03 & 48.92 & 0.0052 \\
1024\hphantom{00} & \hphantom{0}256 & \hphantom{0}90000& 189.31 & 55.89 & 0.0066 \\
1024\hphantom{00} & \hphantom{0}256 & \hphantom{0}120000& 256.55 & 52.04 & 0.0070 \\
2048\hphantom{00} & \hphantom{0}256 & \hphantom{0}60000& 341.09 & 232.16 & 0.0048 \\
\hline
%0.1\hphantom{00} & \hphantom{0}876.0 & \hphantom{0}875.74 & 0.03 \\
%0.01\hphantom{0} & 2441.0 & 2441.0\hphantom{0} & 0.0\hphantom{0} \\
%0.001 & 4130.0 & 4129.3\hphantom{0} & 0.16\\ \botrule
\end{tabular} }
%\begin{tabnote}
\end{center}
%\caption{Derivative relations for Hartley and Fourier transforms}
\end{table}
%\end{doublespace}
The average computation times of 50 realizations given in the Table~\ref{tabtwo} below are in the units of seconds. Again, the times are measured on a Dell Vostro 1700 laptop with dual core of 1.8 GHz and 1GB RAM which is used to run the MATLAB code. As it can be seen on the table, for a very small number of time steps the compressive spectral method does not provide any improvement in the computational effort, like in the case of solitary wave simulations. Again this is mainly due to the computational effort required by the ${l_1}$ minimization. However as the number of time steps gets bigger, the computational effort significantly reduces while the differences in the wave profiles are of negligible importance. Therefore compressive spectral method provides a great computational efficiency compared to the classical spectral method for the water wave simulations.

\section{Conclusion and Future Work}
In this study the compressive spectral method for the simulation of the water waves is introduced. The sparsity property of the water waves, that is with majority of their components being zero, is considered either in time or the frequency domain. For the sparse water wave model in the time domain the well-known solitary wave solutions of the Korteweg-deVries equation are simulated whereas for the sparse water wave model in the frequency domain the well-known Airy (linear) ocean waves with Jonswap spectrum are simulated. It is shown that by using a smaller number of spectral components and the compressive sampling technique it is possible to reconstruct the wavefield with negligible difference compared to the classical spectral method which uses a high number of spectral components. Proposed compressive spectral method is shown to decrease the computational effort significantly by reducing the computation times, especially for large time evolutions.

In tsunami modeling various forms of solitary and asymmetric solitary waves are used. In computationally intensive tsunami models, the compressive spectral method developed in this paper can be used for increasing the efficiency as a future work. Therefore this approach can lead to more detailed and/or computationally more efficient simulations. Additionally, the compressive spectral method offered in this paper can be extended to simulate the nonlinear ocean waves. The sequential, parallel or distributed algorithms can be used for this purpose. Moreover, the compressive sampling methodology can also be incorporated for other type of spectral methods such as those where the Chebyshev, Legendre and other forms of basis functions are used for simulations in various types of domains. Therefore offered methodology can be used for various types of computational schemes  in the broader areas of the applied mathematics and physics.

\end{document}